\documentclass[10pt,a4paper]{article}
\usepackage[utf8]{inputenc}
\usepackage{amsmath}
\usepackage{amsfonts}
\usepackage{amssymb}
\usepackage[round]{natbib}
\usepackage{standalone}
\usepackage{color}
\usepackage{graphicx}
\usepackage{import}
\usepackage{caption}
\usepackage{float}
\usepackage{authblk}

\usepackage{algorithmicx}
\usepackage{algorithm}
\usepackage{algpseudocode}

\title{A Regionalisation Approach for Rainfall based on Extremal Dependence}

\author{K.R. Saunders%
  \thanks{Electronic address: \texttt{K.R.Saunders@tudelft.nl}; Corresponding author}}
\affil{Delft Institute of Applied Mathematics, Delft University of Technology, Delft, Netherlands}

\author{A.G. Stephenson} %
\affil{Data61, CSIRO, Clayton, Victoria, Australia}

\author{D.J. Karoly} %
\affil{School of Earth Sciences, The University of Melbourne, Parkville, Victoria, Australia}
\affil{NESP Earth Systems and Climate Change Hub, CSIRO, Aspendale, Victoria, Australia}

\date{\today}

\begin{document}
\maketitle


\begin{abstract}

To mitigate the risk posed by extreme rainfall events, we require statistical models that reliably capture extremes in continuous space with dependence. However, assuming a stationary dependence structure in such models is often erroneous, particularly over large geographical domains. Furthermore, there are limitations on the ability to fit existing models, such as max-stable processes, to a large number of locations. To address these modelling challenges, we present a regionalisation method that partitions stations into regions of similar extremal dependence using clustering. To demonstrate our regionalisation approach, we consider a study region of Australia and discuss the results with respect to known climate and topographic features. To visualise and evaluate the effectiveness of the partitioning, we fit max-stable models to each of the regions. This work serves as a prelude to how one might consider undertaking a project where spatial dependence is non-stationary and is modelled on a large geographical scale.

\end{abstract}


\section{Introduction}

\begin{figure*}[h!]
\centering
  \includegraphics[width = 0.75\textwidth]{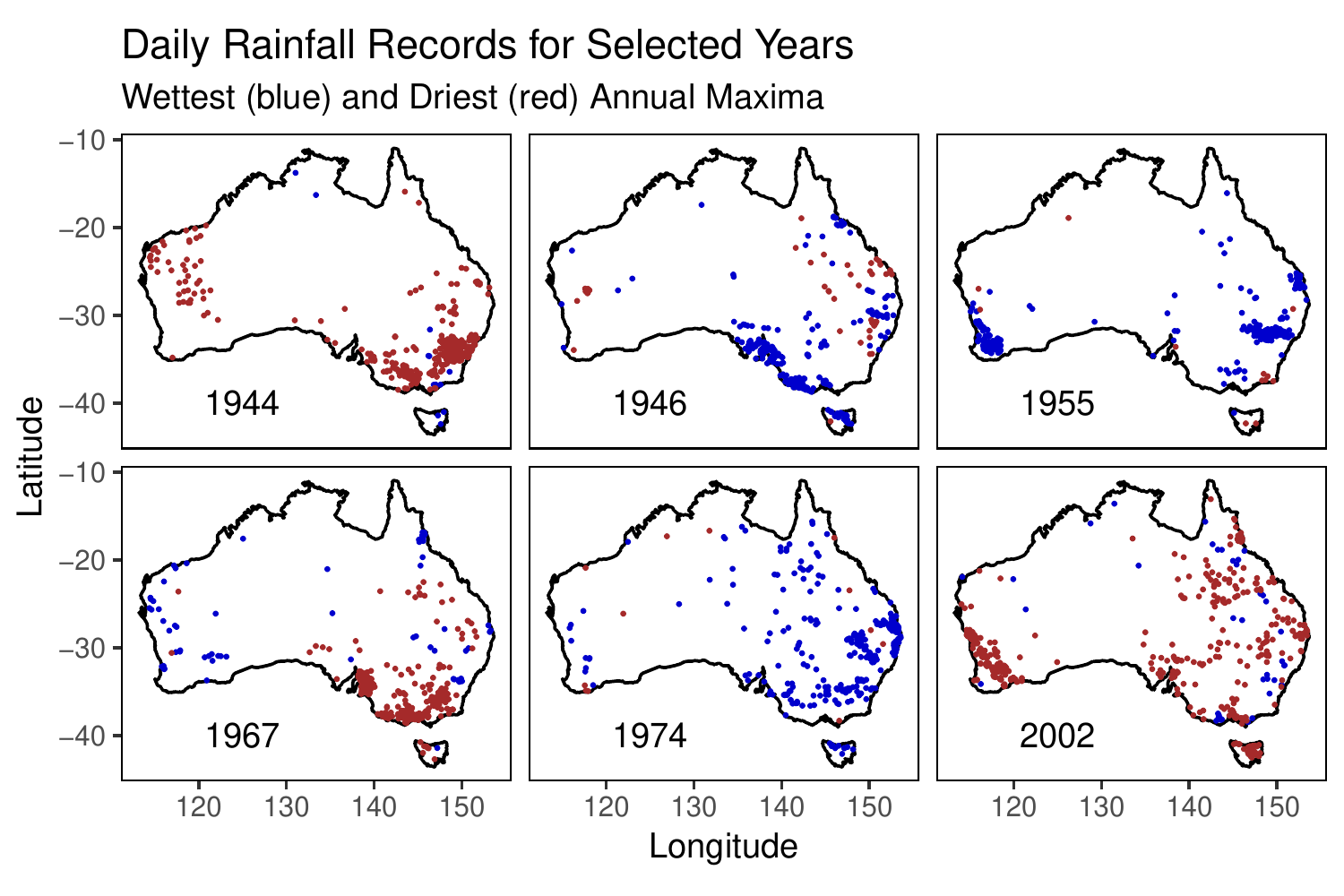}
  \caption[]{For the given year, the plot shows the locations of stations at which the wettest annual maximum was observed (blue) and the driest (red). The years selected are the top three wettest (1946, 1955 and 1974) and top three driest (1944, 1967 and 2002) by proportion of stations. Note that observational periods do vary between stations. Stations often appeared to be clustered tightly in a given colour, but that many clusters can occur in the same year and across large geographical scales.}
  \label{fig:record_summary}
\end{figure*}

The impacts of extreme rainfall and associated flooding can be observed on a scale that covers hundreds of kilometres. For example, the 2011 floods in Australia affected an area the size of France and Germany \citep{queensland2012queensland}. Flooding on this scale is also not unprecedented, with further evidence that extreme rainfall and associated flooding can occur across large geographical scales given in Figure \ref{fig:record_summary}. These historical instances establish the need to understand the spatial range of potential impacts from extreme rainfall. However, for many countries this understanding is lacking, particularly on daily and sub-daily scales.\\  

Statistical models can be used to assess the spatial range of dependence between rainfall extremes, with a summary of some common statistical methods given in \citet{davison2012statistical}. Of particular interest are max-stable processes, which provide a natural extension of univariate extreme value theory to extremes in continuous space with dependence \citep{schlather2002models, de1984spectral}. Modelling rainfall extremes in continuous space is desirable as the risk at locations without stations can be assessed. Max-stable processes also have strong mathematical justification for extrapolating outside the range of the observed data. Given this, these processes have been used in several studies of extreme rainfall; including \citep{saunders2017spatial, dombry2013regular,de2016high}. \\

However, the parametric dependence structure of the max-stable process is often assumed fixed across a given domain for computational and mathematical simplicity \citep{oesting2017statistical}. Depending on the domain, a fixed dependence structure may not be a reasonable modelling assumption. For a large geographical domain however, this assumption is likely to be poor. 
 For example, Australia is one of the largest countries by area, with a diverse climate and complex topographic features \citep{risbey2009remote,stern2000objective}. Assuming a fixed parametric dependence structure is unlikely to yield meaningful results. This presents an obstacle to creating a parsimonious statistical model and reliably identifying which regions are likely to experience similar impacts from extreme rainfall.\\

Promising extreme-value approaches are emerging that model non-stationarity within the dependence as a function of covariates \citep{huser2016non, castro2018time, camilo2017local}. However, these methods are mathematically and computationally complex. As such they are prohibitive for many applied researchers in climatology and hydrology. To understand the how the spatial range of dependence varies for rainfall extremes, a solution is therefore desired in which the method can be quickly implemented and in which the results lead to a simple interpretation.\\

To address this, we present a method for creating regionalisations of rainfall extremes, in which the regions are identified based on extremal dependence. Variations in the size and shape of these regions will indicate the spatial range of the dependence and whether the dependence behaviour is anisotropic. This knowledge can then be translated into insights for assessing and mitigating the potential impacts of extreme rainfall. \\

Regionalisations are common in flood frequency analysis and studies of hydrological extremes. Examples of different approaches to regionalisation based on extreme rainfall are given in \citet{hosking2005regional, carreau2016characterization} and \citet{asadi2018optimal}. For Australia, a regionalisation specific to rainfall extremes does not exist. However, there are regionalisations formed using topography and mean climate \citep{stern2000objective,bureau2015climate}.\\ 

The regionalisation presented here is based on the clustering method presented in \citet{bernard2013clustering}. In this method, a rank-based distance measure is used to cluster stations. This distance measure is related to bivariate extremal dependence via the F-madogram \citep{cooley2006variograms}. Using a rank-based distance is powerful, as no information about climate or topography is required to form spatially homogeneous clusters. This circumvents the challenge of variable selection. Additionally, we are free from distributional assumptions as the F-madogram can be estimated non-parametrically from raw maxima.\\

Where this paper extends the work of \citet{bernard2013clustering} is in the choice of unsupervised learning algorithm. In the original application, $K$-medoids was used for clustering. However, $K$-medoids is sensitive to point density. Additionally if there are too few clusters, $K$-medoids produces spurious clusters when used with the F-madogram distance. We demonstrate these undesirable features using simple examples. For station networks with varying point density, such as Australia, $K$-medoids is therefore ill-suited.\\

We propose using hierarchical clustering instead with the F-madogram distance. This ensures the clusters obtained are not affected by station density and are well informed by extremal dependence. The hierarchical nature of the algorithm also has an interpretation in terms of the changing strength of dependence. We demonstrate how the different clustering methods perform using daily rainfall stations in Australia. We show the serious consequences of incorrectly using $K$-medoids comparing with the results from the more robust hierarchical clustering. We also 
perform an additional classification step. This step converts the clusters from F-madogram space into a euclidean space, giving a more intuitive spatial interpretation. \\

The resulting regionalisation generates valuable insights into the dependence of Australian rainfall extremes. We demonstrate this through a range of examples, highlighting features of climate and topography. We also show how the regions defined using a measure of partial dependence translate to the full dependence of spatial extremes. We achieve this by fitting max-stable models to the stations in each region. The results improve our understanding the spatial range of extreme rainfall events, and how this range varies with increasing dependence strength.


\section{Data}

\begin{figure}[h]
  \centering
    \includegraphics[width=0.8\textwidth]{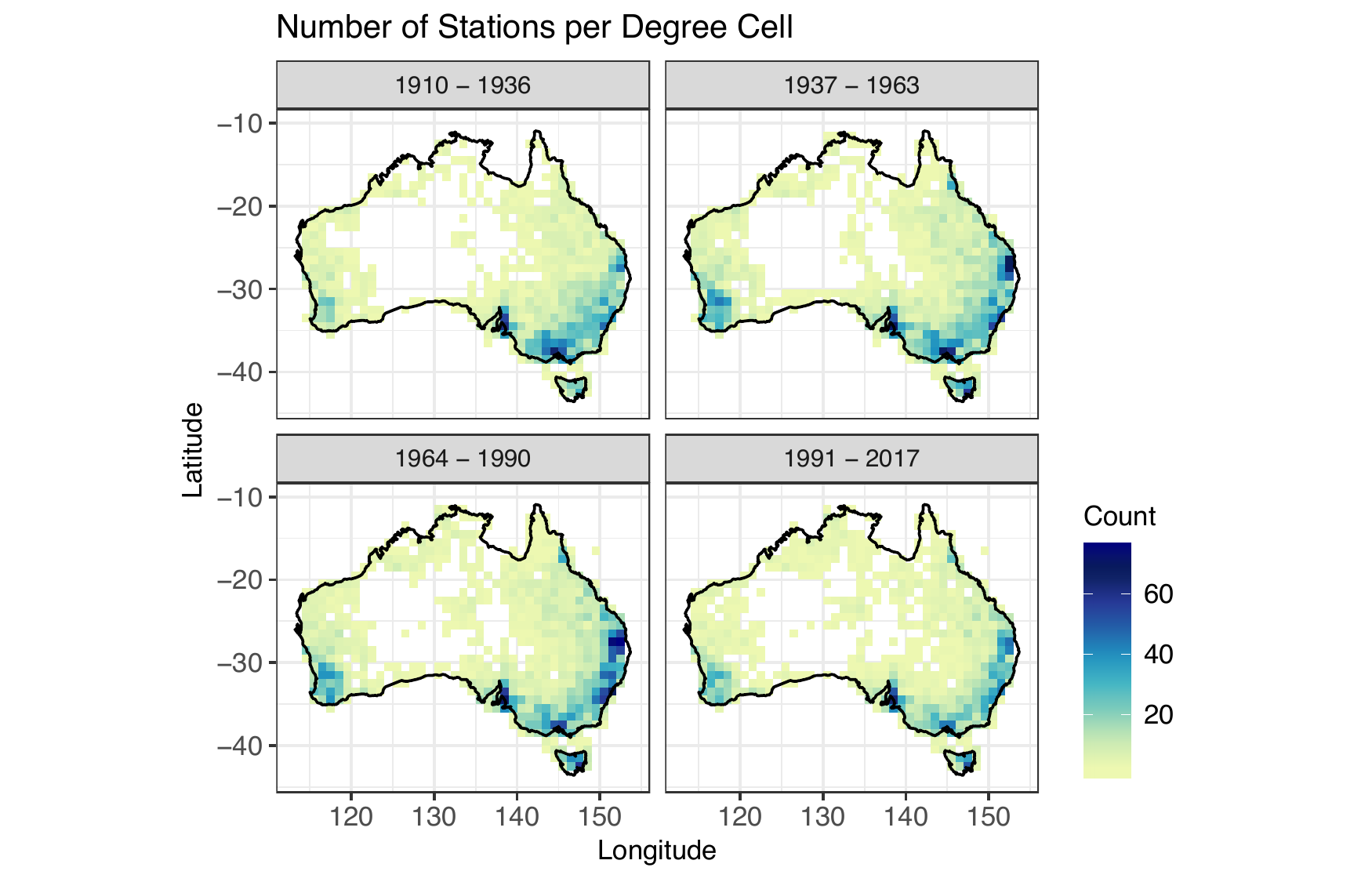}
    \caption[Plot showing the network density of Australian Rainfall Stations]{The plot shows the number of stations within each one degree grid cell that have observations spanning the given time period.}
    \label{fig:stn_binned}
\end{figure}

In this paper, we use the network of daily rainfall stations in Australia. These stations are mainly located near large cities and along the Eastern Australian coast, Figure \ref{fig:stn_binned}. In inland and more remote areas, there are far fewer stations. The station data is obtained from the quality controlled GHCN-Daily dataset \citep{durre2008strategies,durre2010comprehensive} and can be accessed via the R package, rnoaa  \citep{rnoaa}. However, we acknowledge the quality control, while thorough, is of a general design and is not targeted at identifying errors amongst extremal observations \citep{saunders2018investigation}. For example, caution should be exercised when excluding observations flagged as outliers, as these observations may be extremes. \\

The Australian observations within GHCN-Daily are available via a reciprocal agreement with the Australian Bureau of Meteorology. The period we consider is restricted from 1910 to 2017. Prior to 1910 recording practices were not standardised throughout Australia. \\

The analysis is performed using the observed annual maximum rainfall. In extreme value approaches, this is referred to as block maxima. This is in contrast to peaks over threshold, where we are unconcerned with the date of the maxima within the yearly block. To ensure the quality of the observed maxima, we have restricted the data by only considering years which are 90\% complete and stations at which there is a minimum of 20 years of observed maxima. This is necessary to ensure the quality of any extreme--value assumptions \citep[eg.][]{coles2001introduction} and to limit the affects of missing maxima \citep[eg.][]{haylock2000hqdataset}. 

\section{Clustering Method}

In the following section, we outline how to perform the clustering for the regionalistion. This includes describing the choice of the dissimilarity and choosing an appropriate clustering algorithm. 


\subsection{Clustering Dissimilarity}

A notion of dissimilarity (or similarity) between two points is required to apply clustering algorithms, with the type of dissimilarity chosen determining the cluster structure. For this application, following \citet{bernard2013clustering}, we have chosen to use the F-madogram distance \citep{cooley2006variograms}\footnote{The dissimilarity used in clustering can be a distance, but it does not necessarily need to satisfy the triangle inequality \citep{friedman2001elements}.}. The F-madogram distance has an interprettion in terms of the pairwise dependence strength of extremes. The resulting cluster structure therefore inherits a meaningful, physical interpretation. 

\subsubsection{F-madogram}

The F-madogram \citep{cooley2006variograms} links ideas of dependence in spatial statistics and dependence in extreme value theory. In spatial statistics a variogram \citep[eg.][]{cressie2015statistics} is commonly used to understand the dependence between two locations in a stochastic process. However, for extremes the variogram is often undefined, as the distributions can be heavy-tailed and the variance is not finite. In contrast, the F-madogram, which is conceptually similar, is defined for heavy-tailed distributions.\\

Let $\mathcal{S}$ be the set of $n$ stations to be clustered. For $x_i \in \mathcal{S}$, define $M_i$ as the random variable representing the annual maximum, daily rainfall at that station. Let the distribution function associated with $M_i$ be $F(z)$. We can estimate $F(z)$ empirically
\begin{align}
	\hat{F}_i(z) = \dfrac{1}{|\mathcal{Y}_i|}\sum_{y \in \mathcal{Y}_i}\mathbb{I}\left(M_i^{(y)} < z\right),
\end{align}
where $\mathcal{Y}_{i}$ is the set of years for which there are annual maximum observations at $x_i$. 
For stations $x_i \in \mathcal{S}$ and $x_j \in \mathcal{S}$, the F-madogram is given by the mean absolute difference (MAD) between two distribution functions and can be estimated non-parametrically using
\begin{align}\label{eqn:fmado_dist}
	\widehat{d}(x_i, x_j) = \dfrac{1}{2 |\mathcal{Y}_{ij}|} \sum_{y \in \mathcal{Y}_{ij}} \left| \hat{F}_i\left(M_i^{(y)}\right) - \hat{F}_j\left(M_j^{(y)} \right)\right|,
\end{align}
where $\mathcal{Y}_{ij}$ is the set of years when both stations $x_i$ and $x_j$ have annual maximum observations. Note that $\mathcal{Y}_i$ and $\mathcal{Y}_{ij}$ may differ depending on missing observations. \\

Non-parametric estimation of the F-madogram avoids distributional assumptions and model fitting. This makes using the this distance for clustering particularly powerful, as no external information about climate or topography is required and there is no need for variable selection. However, this assumes that the annual maxima are stationary in time. It may be necessary to remove trends depending on the application, such as in the case of temperature extremes \citep{bador2015spatial}.

\subsubsection{Bivariate Extreme Value Distribution}

The link between the F-madogram and extreme value theory provides the cluster structure with a physical interpretation in terms of the dependence of extremes. For any pair of stations, $x_i$ and $x_j$, if the distribution of $(M_i,\, M_j)$ is well approximated by a bivariate extreme value distribution then  
\begin{align}
	\mathbb{P}\left( M_i \leq z_i,\, M_j \leq z_j \right) = \exp\left\{ - V_{ij} \left( \dfrac{-1}{\log F_i(z_i)}, \, \dfrac{-1}{\log F_j(z_j)}\right) \right\},
	\label{eqn:bivariate_distbn}
\end{align}
where the exponent measure $V_{ij}(a,b)$ is given by 
\begin{align}
	V_{ij}(a, b) = 2 \int_0^1 \max \left(\dfrac{w}{a}, \dfrac{1-w}{b} \right) \text{d}H_{ij}(w),
	\label{eqn:bi_extremal_component}
\end{align}
and $H_{ij}$ is any distribution function on $[0,1]$ with expectation equal to 0.5 \citep[eg.][]{resnick1987extreme, de2006extreme}.\\

In the special case where $z_i = z_j = z$, the bivariate extreme value distribution of equation \eqref{eqn:bivariate_distbn} reduces to 
\begin{align}
	 \mathbb{P}(M_i \leq z, M_j \leq z) &= \left[\mathbb{P}(M_i \leq z)\mathbb{P}(M_j \leq z)\right]^{V_{ij} (1,1) / 2},
\end{align} 
where 
\begin{align} 
V_{ij} (1,1) &= \theta(h)
\end{align} 
and $\theta(h)$ is the extremal coefficient, with $h = x_j - x_i$ \citep[eg.][]{naveau2009modelling}. The range of $\theta(h)$ is $[1, 2]$, where the lower bound of the interval corresponds to dependence of $M_i$ and $M_j$, and the upper bound conversely indicates independence. The value of $\theta(h)$ therefore provides an indication of the partial dependence between the maxima at the two locations $x_i$ and $x_j$ when $z_i = z_j = z$.\\

The F-madogram dissimilarity can be expressed as a function of the extremal coefficient \citep{cooley2006variograms} 
\begin{align}
	d(x_i, x_j) &= \dfrac{\theta(h) - 1}{2(\theta(h) + 1)},
\end{align}
where the range of $d(x_i, x_j)$ is $[0, \tfrac{1}{6}]$. Therefore when it is suitable to approximate the pairwise distribution of annual maxima with bivariate extreme value distributions, clusters formed using the F-madogram distance will have an interpretation in terms of partial dependence of extremes.\\

Equally, we could have used $\theta(h)$ for the clustering dissimilarity. However, the F-madogram as a mathematical object can be estimated independently of distributional assumptions and therefore of extreme value assumptions. As such, it offers a more flexible choice for the dissimilarity.

\subsubsection{Practicalities of missing dissimilarities}

All pairwise dissimilarities are required for clustering. However, unlike gridded datasets, observational periods at two stations may not overlap due to missing data. Additionally, if the number of overlapping years is small, the F-madogram distance cannot be estimated reliably. 
Therefore to maximise the station data available, particularly in sparse regions, missing distances were interpolated. \\

At large euclidean distances, we expect the maximum rainfall observed at pairs of stations to be close to independent. Given this, these missing dissimilarities were interpolated as $\tfrac{1}{6}$. This is a reasonable assumption and greatly reduces the missing dissimilarities. Also, for a station that has been renamed, the euclidean distance between stations may be 0 and then the missing F-madogram distance is interpolated as 0.\\

For the remaining missing dissimilarities we we fit regional linear models to the logarithm of Euclidean distance. From this model we predicted missing distances and while these predictions do not approximate local dependence well, they do serve as a reasonable approximation of overall dependence. At very small euclidean distances predictions could take negative values, so the maximum of the predicted F-madogram distance and zero was taken. 

\subsection{Clustering Algorithm}

In the previous section, we provided the necessary information about estimating the F-madogram distances and understanding the physical meaning behind the clustering structure. In this section, we discuss the choice of clustering algorithm. We contrast cluster structures generated using $K$-medoids and hierarchical clustering, highlighting subtle features of these different algorithms. In particular, we discuss the suitability of these algorithms for our application. 


\subsubsection{K-medoids}\label{app:Kmedoids}

In clustering application of \citet{bernard2013clustering}, $K$-medoids clustering was applied with the F-madogram distance. In $K$-medoids, the goal is to find $K$ clusters such that the sum of dissimilarities relative to a representative point within each cluster is minimised. This representative point is known as the medoid. Denote the medoids $\{m_k \,|\, k = 1 \dots, K \}$ and their associated clusters $\{C_k \,|\, k = 1 \dots, K \}$, where $K \leq n$. To partition the points we can use the PAM algorithm \citep{kaufman1990partitioning}, see Table \ref{algorithm:pam}. \\

\begin{algorithm}

\begin{algorithmic}[1]
\Procedure{Partitioning Around Medoids}{} \smallskip

\State Choose the number of clusters, $K$ \smallskip


\State Randomly select $K$ points in $\mathcal{S}$ as the initial medoids \smallskip



\State Determine the closest medoid to each point \smallskip 

\State Cluster points that share the same closest medoid \smallskip 

\For {$k$ in $1, \dots, K$} \smallskip

	\State Find the point within that cluster, $C_k$, such that 
	\begin{align}\label{eqn:medoid}
		\quad\quad\quad m_k^*& = \mathop{\mathrm{argmin}}\limits_{x_i \in C_k} \sum_{x_j \in C_k} \hat{d}(x_i, x_j). 
	\end{align}\smallskip
	\quad\quad\quad \nonumber{This point minimises the sum of }\\
	\quad\quad\quad \nonumber{dissimilarities within that cluster.}\smallskip
	\If {$m_k^* \neq m_k$} \smallskip

    	\State Update the medoid so that $m_k = m_k^*$ \smallskip

	\EndIf \smallskip

\EndFor \smallskip

\If {Any of the medoids were updated} \smallskip

    \State Repeat steps 4. -- 12.  \smallskip

\EndIf
\EndProcedure
\end{algorithmic}
\caption{K-medoids clustering}\label{algorithm:pam}
\end{algorithm}

Like many clustering algorithms, PAM converges to a local minimum, but not necessarily the global minimum. It is therefore advisable to repeat PAM with different initialisations of medoids to help ensure the consistency within the performance of the algorithm. We do not discuss how to select $K$ optimally for $K$-medoids here, but implementations of various methods can be found in \citet{NbClustRPackage}.

\subsubsection{Implicit Assumptions}

Within unsupervised learning there is no true structure. However, we often still have implicit assumptions about the structure form. For our application, we have the expectation that two stations that are far away in euclidean space will be clustered differently as the extremes at these station are independent. We also have the expectation that stations that are geographically close will be clustered together as they are likely highly dependent.\\

\begin{figure}[!htb]
	\center
		\includegraphics[width=0.7\textwidth]{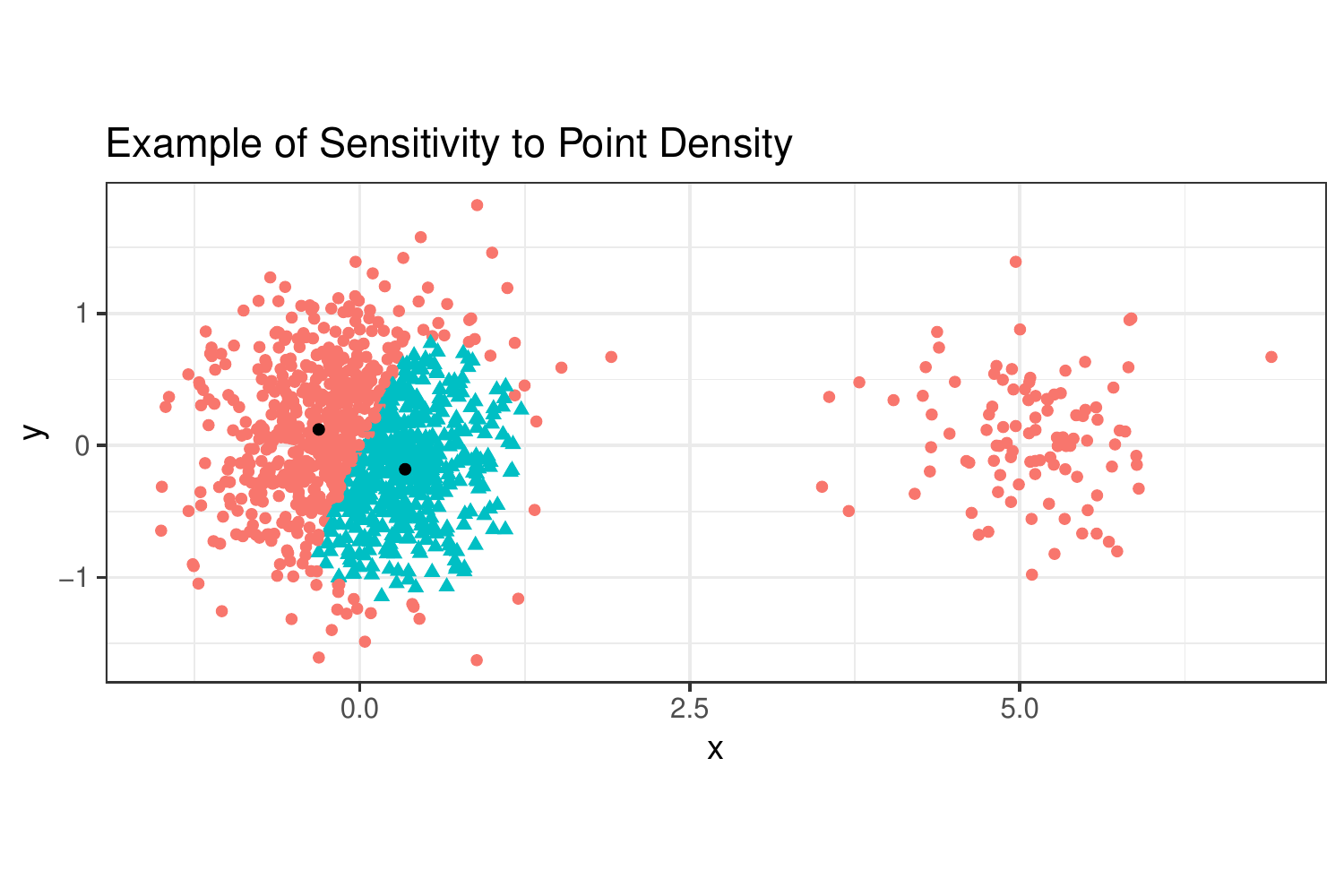}
     \caption{Example of K-medoids clustering showing that the assignment of points to clusters is sensitive to the spatial density of points. It is not until the value of $K$ is increased to 4 or more that the second group is identified, and this is not an optimal assignment of points to medoids.}
     \label{fig:EgDensity}
\end{figure}

\begin{figure}[!h]
	\center
		\includegraphics[width=0.7\textwidth]{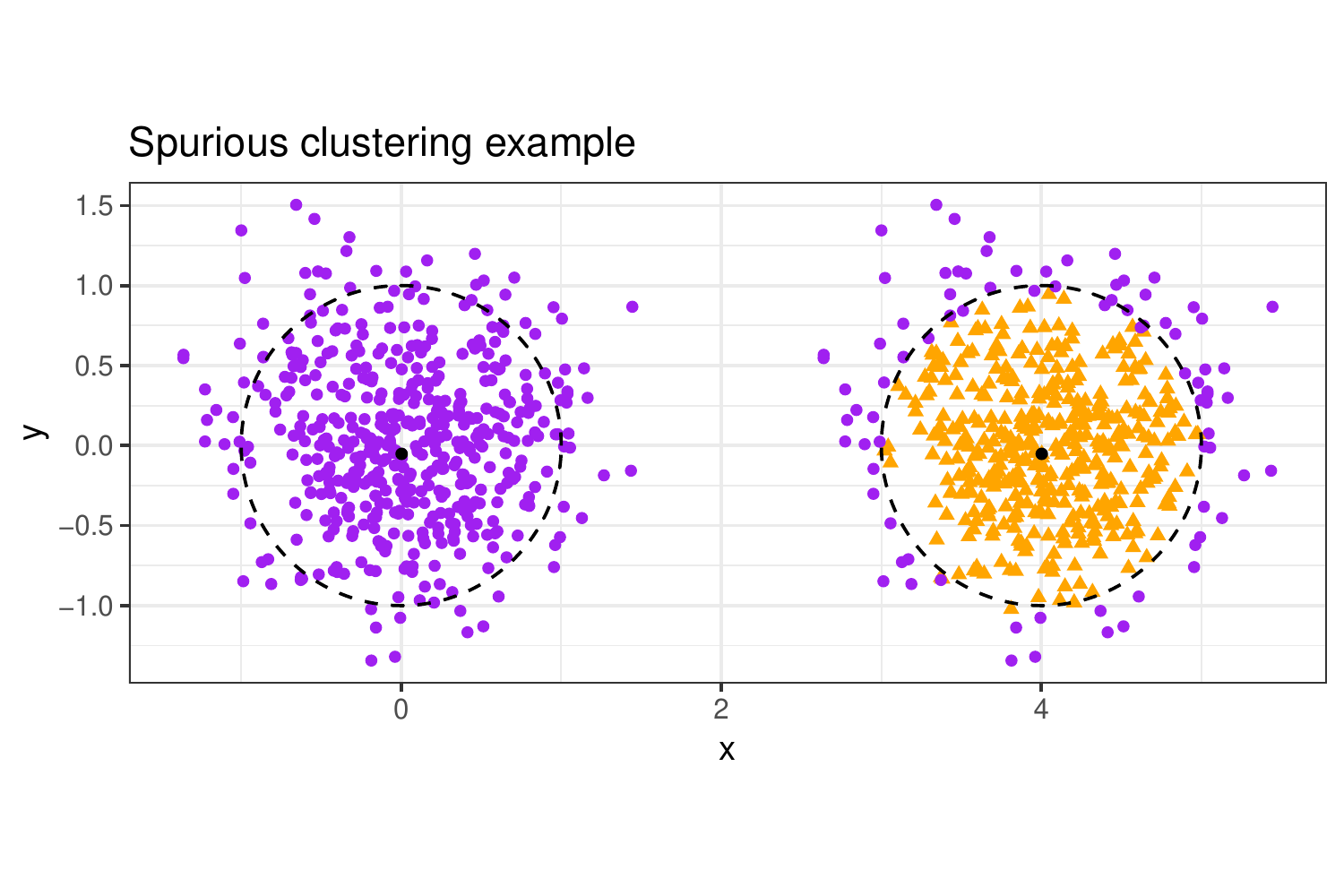}
     	\caption{Example of K-medoids clustering showing undesirable clustering behaviour when points are equidistant from all medoids.}
     	\label{fig:EgSpurious}
\end{figure}

Consider the two examples shown in Figure \ref{fig:EgDensity} and Figure \ref{fig:EgSpurious}. In each of these examples the structure is known and there are two groups of points. However, $K$-medoids clustering does not recover the two groups correctly. We have not used the F-madogram distance in these examples. Instead, it is more intuitive to think in euclidean space, so the distance used is 
\begin{align}
	d(x_i, x_j) = \max \left( \|x_j - x_i\|, 1 \right),
\end{align}
where $\| \cdot \|$ is the euclidean distance. However, we restrict the maximum value this distance can take to 1, in order to mimic the finite range of the F-madogram distance. \\ 

The example in Figure \ref{fig:EgDensity} shows that $K$-medoids clustering is sensitive to the spatial density of points. The location of the medoids, the representative object within each cluster, is biased toward regions of higher spatial point density. As such, points in the smaller group are clustered in an undesirable way. Under the optimisation this is not unexpected. However, clustering such as this is not in keeping with our implicit assumptions. Also as the resulting clusters are not robust to the spatial density of points, any interpretation of the structure in terms of extremal dependence will not be meaningful. Gridded datasets to some extent would be immune to this problem, provided proper consideration is given to land-sea and domain boundaries. \\

More seriously however, is the example of Figure \ref{fig:EgSpurious} that shows that $K$-medoids can produce spurious clustering. Here, we have drawn a circle of radius one around each medoid, where points outside of these circles are of distance 1 to either medoid. Under the optimisation, these points can be assigned randomly to either cluster without penalty.  
More insidiously however, is that all these points are labeled the same. This is due to a numeric ordering within the standard algorithm. Groups of points can therefore appear to be clustered meaningfully, even though they are not. Consequently, if there are two few medoids, points will be assigned randomly. Additionally, given that the location of medoids is biased toward regions of higher point density, under $K$-medoids, sparsely located points will always be clustered spuriously.\\

These examples demonstrate that the selection of the clustering method needs to be evaluated relative to the dataset to ensure the clustering is meaningful. Given that the Australian station network is highly variable in terms of spatial point density, it is highly unlikely that a cluster structure obtained using $K$-medoids and the F-madogram distance will be informative in terms of extremal dependence. As such an alternative method is needed for clustering.\\

 Two of the other most common methods are $K$-means and hierarchical clustering. $K$-means however is subject to the same failings demonstrated in Figures \ref{fig:EgDensity} and \ref{fig:EgSpurious}. Regardless however even for suitable datasets, $K$-means is not an appropriate choice given euclidean assumptions and a standard algorithm implementation in terms of points not distances \citep{friedman2001elements}. Hierarchical clustering in contrast, can be used with an F-madogram distance to produce meaningful structures in terms of extremal dependence. 

\subsubsection{Hierarchical Clustering}

In hierarchical clustering an ordered sequence of partitions is created. This hierarchy of partitions has a natural intuition for our application, and can be interpreted as partitions of points based on strong dependence to weaker dependence. Graphically, this ordered sequence of partitions can be represented using a dendrogram. Let each point be its own cluster (leaf). Branches in the dendrogram are formed by successively combining leaves and other branches until all points are grouped together. For each merge, a new partition of the points is induced. The successive merging of branches therefore creates the ordered partition of points.\\

To decide how branches should be merged the definition of distance needs to be extended from between two points to include the distance between two groups of points. This is known as the linkage criterion \citep{murtagh1983survey, mullner2011modern, murtagh2014ward}. Let $C_k$ and $C_{k'}$ be two different clusters of points. We use the average linkage criterion  
\begin{align}\label{eqn:upgma}
	d(C_k, C_{k'}) = \frac{1}{|C_k|\,|C_{k'}|} \sum_{x_k \in C_k} \sum_{x_{k'} \in C_{k'}} d(x_k, x_{k'}).
\end{align}
Using the linkage criterion, we can construct an agglomerative dendrogram using the algorithm in Table \ref{algorithm:hclust}.

\begin{algorithm}

\caption{Hierarchical clustering}\label{algorithm:hclust}
\begin{algorithmic}[1]
\Procedure{agglomerative}{} \smallskip

\State Let each point form its own cluster \smallskip

\State Merge the clusters with the smallest dissimilarity \smallskip 

\State Update the dissimilarities relative to the new cluster \smallskip\quad\quad according to the linkage criterion \smallskip

\State Repeat steps 2--4, until all points are combined in a \smallskip\quad\quad single cluster \smallskip

\EndProcedure
\end{algorithmic}
\end{algorithm}

To determine an assignment of points into clusters, we need to select one of the partitions generated by the dendrogram. This can be done by cutting across the tree at a height $h$, and assigning the points in same branch to the same the cluster. Equivalently, we can specify the number of clusters, $K$, and choose the cut height that corresponds to this number of clusters.\\

The height of the cut should be made with reference to the desired strength of association between the clusters, with the height at which the branches are fused determining the strength of association between two clusters. Therefore for two branches joined at the bottom of the tree, this suggests the points in these branches are strongly associated. For branches joined at the top of the tree, this suggests a much weaker association between the groups of points. Standard methods for choosing the cut height include the gap statistic, see \citet{tibshirani2001estimating}. Equally valid, is choosing a cut height based on user knowledge. We do this based on visualising the extremal dependence.\\

In hierarchical clustering different linkage criterion will induce different dendrograms and consequently different clusters. The average linkage criterion successfully recovers the two groups show in Figures \ref{fig:EgDensity} and \ref{fig:EgSpurious}. However, this is not the case for many standard linkage rules. Therefore a caveat of this method is that caution is needed in selecting an appropriate linkage criterion for the application.


\section{Classification}
\label{ssub:classification}

Hierarchical clustering is performed in F-madogram space, however for most applications regions are needed in euclidean space. As such, an additional classification step is needed. This step is also necessary to classify locations without a station and to identify boundaries between two clusters for predictive purposes.\\

We have used a weighted $k$-nearest neighbour classifier ($wk$-NN) \citep{dudani1976distance} to classify a grid points covering our domain and to convert the clustering to a regionalisation. We chose the $wk$-NN method as it is non-parametric, based on minimal assumptions, and can form non-linear boundaries. 
In standard $k$-nearest neighbour classification ($k$-NN) \citep[eg][]{hastie2009unsupervised}, points are classified similarly to the majority of their $k$-nearest neighbours without using weights. However, the relationship between the F-madogram and euclidean distance is not linear, so a weighted classifier is more appropriate for this application. The classification algorithm works as given in Table \ref{algorithm:wknn}. \\

\begin{algorithm}
\begin{algorithmic}[1]
\caption{Classification}\label{algorithm:wknn}

\State The stations, $\mathcal{S}$ from the clustering will form the training points for the classification \smallskip

\State For $x_i \in \mathcal{S}$, define $l(x_i)$ to be the label assigned with $x_i$\smallskip

\State Grid the domain for classification \smallskip

\Procedure{Weighted $k$ nearest neighbours}{} \smallskip

\State Choose the number of nearest neighbours, $k_{nn}$, \smallskip \quad \quad where $k_{nn} \leq n$ \smallskip

\For {each grid point, $g$} \smallskip

\State According to euclidean distance, get the $k_{nn} + 1$ \smallskip \quad \quad  nearest neighbours to $g$ in $\mathcal{S}$ \smallskip

\State Let the furthest of these neighbours be $n_f$ \smallskip 

\State Let the set, $\mathcal{N}$, contain the other nearest \smallskip \quad \quad \quad  neighbours, $\{n_j \,|\, j = 1, \dots k_{nn}\}$ \smallskip

\For {each of the nearest neighbours, $n_j \in \mathcal{N}$} \smallskip

	\State Standardise the euclidean distances between \smallskip \quad \quad \quad \quad \quad $n_j$ and $g$
	\begin{align}
		\quad \quad \quad \quad s(n_j) = \dfrac{\lVert g -  n_j \rVert}{{\lVert g - n_f } \rVert}.
	\end{align}

	\State We used an inverse weighted kernel to weight \smallskip \quad \quad \quad \quad \quad each neighbour.

	\State Get the associated weight for the neighbour, \smallskip \quad \quad \quad \quad \quad $n_j$,

	\begin{align}
	\quad \quad \quad \quad \quad w(n_j) = s(n_j)^{-1}.
	\end{align}
	
\EndFor \smallskip

\State Let $\mathcal{C}$ be the set of labels associated with the \smallskip \quad \quad \quad \quad
neighbours in $\mathcal{N}$ \smallskip

\State Determine the label of the majority of the \smallskip \quad \quad \quad \quad weighted $k_{nn}$ nearest neighbours

	\begin{align}
	\smallskip \quad \quad \quad \quad l^* = \mathop{\mathrm{argmax}}_{l^* \in \mathcal{C}} \left( \sum_{l^* \in \mathcal{C}} \sum_{j=1}^{k_{nn}} w(n_j) \mathbb{I}(l(n_i) = l^*) \right),
	\end{align}

\State Classify $l(g)$ with the majority label, $l^*$ \smallskip

\EndFor \smallskip

\EndProcedure

\end{algorithmic}
\end{algorithm}
 
There is a variance bias trade-off when selecting the number of nearest neighbours, $k_{nn}$. However, when the clusters are well separated in euclidean space there are a large range of suitable $k_{nn}$ values. Considerations for this specific application are that we require $k_{nn}$, such that erroneously clustered stations do not impact the classification, and smaller clusters of only a few stations are not engulfed by a larger cluster and its label. It can be difficult to find an automated metric that will respect this latter criteria. Given the large range of suitable values, through visualisation and user knowledge, we used a value $k_{nn} = 15$.


\section{Visualising Dependence}

\begin{figure*}[h!]
	\centering
  \includegraphics[width = 0.75\textwidth]{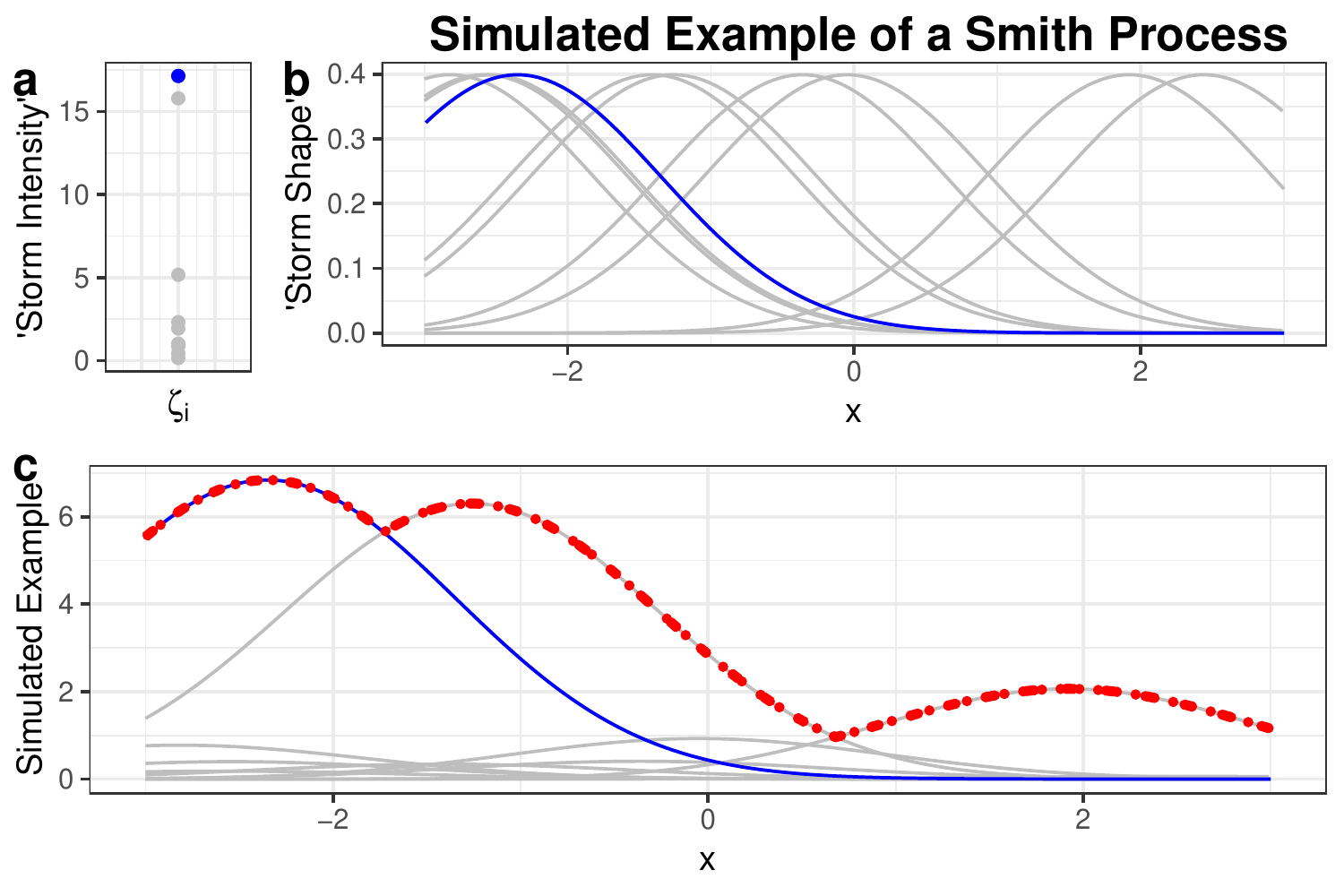}
  \caption[]{This figure shows a visual example of each of the components that comprise the Smith Process in one-dimension. Figure (a) shows points simulated from the inhomogeneous Poisson process, $\{\zeta_i\}$. Figure (b) shows a standard Gaussian density subject to random translations given by $U_i$. Figure (c) shows the product of figures (a) and (b), with an example carried through all figures shown in blue. The resulting simulation of a max-stable process is shown in red, and is given by the pointwise maxima over all scaled `storm-shapes' shown in gray. Here only finitely many simulations of the index $i$ are shown, however, this is all that is necessary to produce a simulated example of the Smith Process (see \citealt{schlather2002models} for details).}
  \label{fig:smith}
\end{figure*}

Part of our motivation for creating this regionalisation was to understand the range of spatial dependence and scale of potential impacts from extreme rainfall. However, the distance used only partially reflects the full extremal dependence. Therefore to consider whether a partitioning forms an appropriate regional summary relative to the full dependence structure, we will fit max-stable processes to the stations in each region.  \\

Max-stable process provide a natural extension from univariate extreme value theory and the  GEV distribution, to models for extremes in continuous space with dependence \citep{de1984spectral, schlather2002models}. The canonical example of these processes is the Smith Model \citep{smith1990max}. This model offers an intuitive storm shape interpretation, where a storm shape is scaled by a storm intensity and the pointwise maxima over infinitely many of these scaled-storms forms a realisation of the max-stable process. Mathematically
\begin{align}
Z(x) \stackrel{d}{=} \max_{i \geq 1} \zeta_i W(x - U_i; 0, \Sigma), \quad x \in \mathcal{X} \subset \mathbb{R}^2, 
\label{eqn:maxstable}
\end{align}
where $\{\zeta_i: i \geq 1 \}$ are points from a Poisson process on $(0, \infty)$ with intensity $\zeta^{-2}\text{d}\zeta$ and $W(\cdot ; 0, \Sigma)$ is a two-dimensional Gaussian density, with mean zero and covariance matrix $\Sigma$. Here, $U_i$ are points of a homogeneous Poisson process defined on $\mathbb{R}^2$ that provide random translations of bivariate Gaussian. A visual representation of this processes in 1-dimension is given in Figure \ref{fig:smith}. The univariate marginals of this max-stable process are assumed to follow a standard Fréchet distribution.\\

The Smith model is used here due to its simplicity and as the dependence structure of this process is Gaussian. We can therefore visualise the dependence in two-dimensional euclidean space using ellipses. The direction and size of these ellipses has a natural interpretation in terms of anisotropy and the range of the dependence. 
For the Gaussian density, the probability of a point, $x$, lying within a radius, $r$, of the mean is given by the Chi-squared distribution with two degrees of freedom
\begin{align}
  \mathbb{P}(\,||x - \mu|| < r) &= 1 - \exp\left(\frac{-r^2}{2}\right). 
\end{align}
For our elliptical curves, we have chosen $r$ to correspond to the $1\%$ level curve, for which $r \approx 3$. However, within the formulation of the Smith model the mean is zero as the Gaussian is subject to random spatial translations. Therefore to centre our the elliptical curves, we use the coordinate of the median longitude and median latitude of all suitable stations in the region, $x_0$. The parameterisation of the ellipses is then given by 
\begin{align}
  x = x_0 + r(\cos \theta, \sin \theta) M,
\end{align}
where $M$ is obtained from the Choleski decomposition of the covariance matrix, $\Sigma = M^TM$.\\

In general, if the partition is a good representative summary then we expect that the ellipses will have minimal overlap. If the ellipses were to overlap, this could indicate that points in the intersection could reasonably have been assigned to either cluster and there may be too many clusters. If we have too few clusters, then more ellipses could be added to summarise dependence. \\

To fit the Smith model we use composite likelihood, see \citet{padoan2010likelihood} for details. In composite likelihood, the sum over bivariate likelihood functions is optimised to obtain parameter estimates. Composite likelihood is used as it is not possible to optimise the full likelihood in higher dimensions where there are large numbers of stations \citep{castruccio2016high, dombry2017full}. As we are primarily interested in the dependence parameters, we first fit the marginals distributions using standard maximum likelihood and standardise our marginals, prior to fitting the dependence parameters using composite likelihood. \\

We acknowledge that the Gaussian storm shape in the Smith model is a crude approximation of physical rainfall and there are other other max-stable processes we could have chosen (see \citealt{dey2015multivariate}). However, as we wish to visualise the full dependence, the the Smith process serves as a useful exploratory tool. Additionally the code for fitting a Smith model with anisotropy is readily available in the SpatialExtremes package \citep{SpatialExtremesPackage}, so the research and method is easily reproducible by others. However, we caution that appropriate starting values are often necessary to ensure convergence of the optimisation routine.


\section{Results}

\subsection{Hierarchical Clustering compared with K-medoids}

\begin{figure*}[!htb]
	\center
		\includegraphics[width=0.75\textwidth]{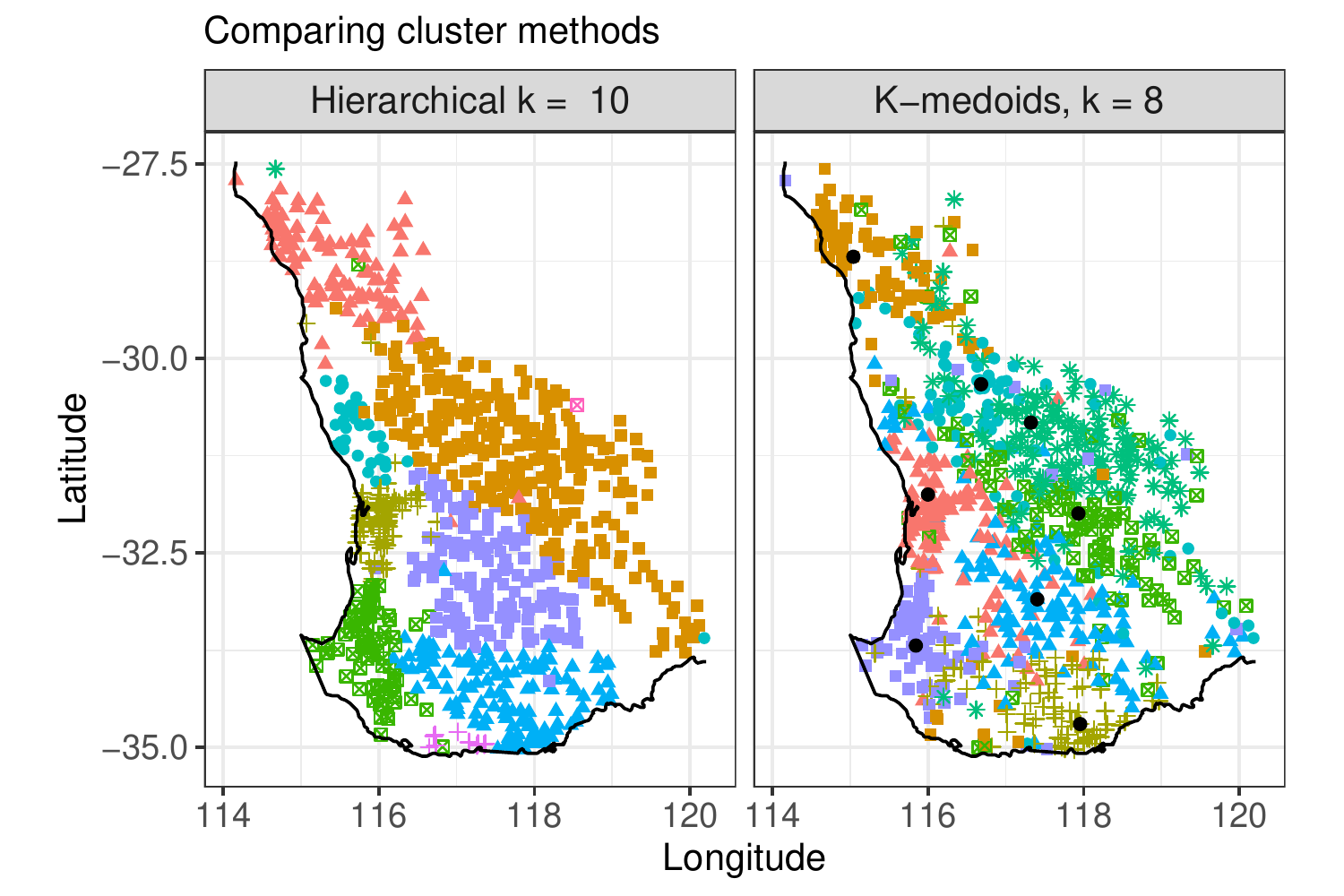}
     \caption{Comparison of hierarchical clustering and K-medoids clustering for a set of stations in Southwest Western Australia.}
     \label{fig:hclust_vs_kmed}
\end{figure*}

To highlight the impact of the choice of clustering algorithm, we have clustered stations in Southwest Western Australia using both hierarchical clustering and $K$-medoids, Figure \ref{fig:hclust_vs_kmed}. Under hierarchical clustering, we observe clearer separation of the clusters in euclidean space. This improved cluster cohesion is a benefit of the hierarchical algorithm having an agglomerative (bottom up) approach.\\

We also note that under hierarchical clustering, clusters can consist of a single station. Therefore to compare the clustering under the two different algorithms, we have chosen realisations where there are 8 core clusters that contain 10 or more stations. The ability of hierarchical clustering to have clusters of smaller size means that groups of stations with weaker dependence are not amalgamated into a larger groups at the expense of the overall cluster cohesion (see Figure \ref{fig:EgDensity}). It also prevents the occurrence of stations being clustered spuriously (see Figure \ref{fig:EgSpurious}). In Figure \ref{fig:hclust_vs_kmed}, we observe the effects of spurious $K$-medoids clustering as there is a large geographical separation between some stations and their respective medoids. For example, at the coordinates (116, -31) and (117, -35). For these reasons, we find hierarchical clustering superior for this application and use this method for the remainder of the paper.

\subsection{Classification} 

\begin{figure}[!htb]
	\center
		\includegraphics[width=0.5\textwidth]{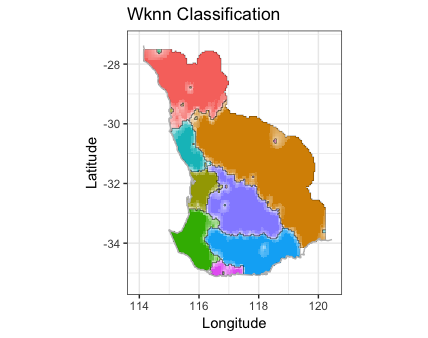}
     \caption{Weighted $k$ nearest neighbour classification showing cluster boundaries from the hierarchical clustering in Figure \ref{fig:hclust_vs_kmed}.}
     \label{fig:classify}
\end{figure}

Figure \ref{fig:classify} shows the classification from the hierarchical clustering for a value of $k_{nn} = 15$. Due to the quality of the original clustering in F-madogram space and separation of clusters in euclidean space there was very little difference for higher $k_{nn}$ values. However, classification does offer the advantage in that we have regional boundaries and do not need to visualise large numbers points.

\subsection{Ordered Partitions} 

\begin{figure*}[!h]
	\center
		\includegraphics[width = \textwidth]{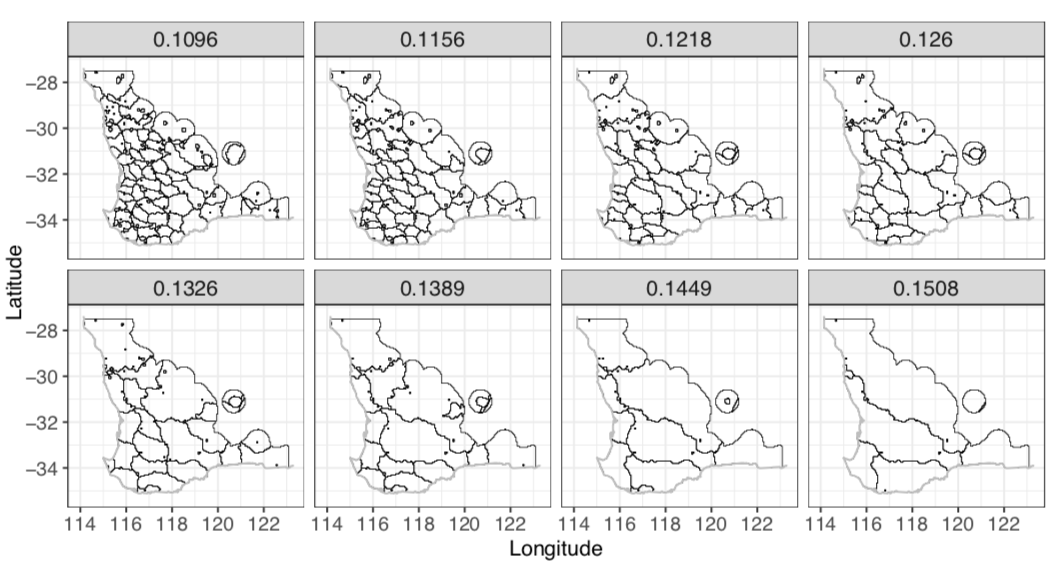}
     \caption{Different regionalisations of Southwest Western Australian created using different cut heights in hierarchical clustering. The cut height is given in the facet label.}
     \label{fig:cluster_summary}
\end{figure*}

We mentioned earlier that one of the benefits of hierarchical clustering is that an ordered sequence of partitions is generated. In Figure \ref{fig:cluster_summary}, we show the evolution of these partitions for a range of cut heights for Southwest Western Australia. We observe that at the lower cuts heights that the regions are small in size. While at higher cut heights, where the dependence between clusters weakens, these smaller regions are amalgamated to form larger regions. Visualising the evolution of these ordered partitions helps our understanding of how the size of these regions changes with increasing strength of extremal dependence.\\

Additionally the size and direction of the regions can then be interpreted relative to known climate or topography. We observe here that coastal clusters are generally smaller indicating that extreme rainfall is being driven by convective rainfall in these areas \citep{risbey2009remote}. Whereas further inland, the size of clusters is larger, particularly as dependence weakens, and orientation of these clusters is consistent with the movement of frontal systems \citep{risbey2009remote}.

\subsection{Meaningful Cut Heights}

\begin{figure}[!htb]
	\center
		\includegraphics[width=0.7\textwidth]{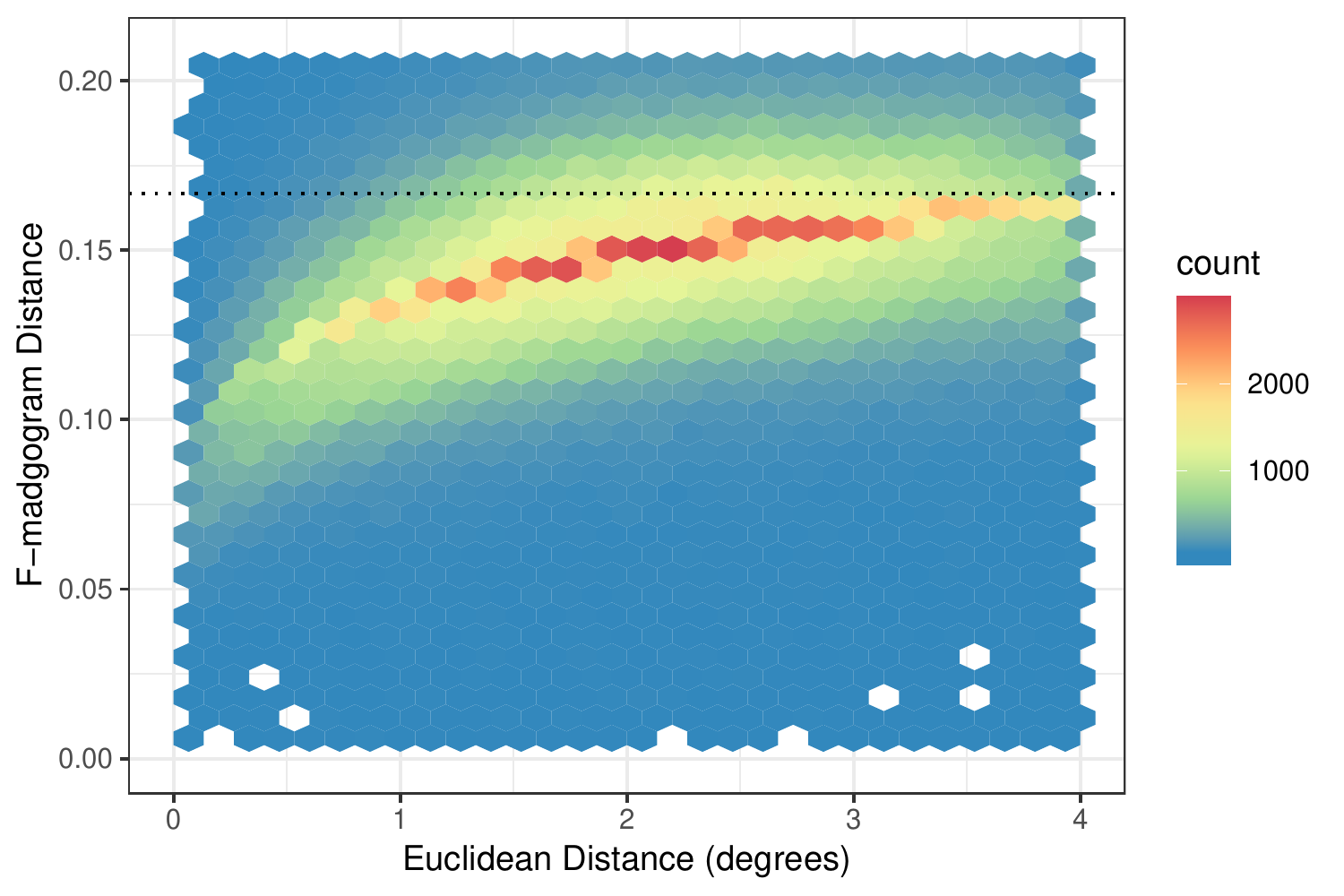}
     \caption{Plot of the F-madogram distance relative to euclidean distance. Given the number of pairs we have binned the data instead of showing a scatter plot. Note that the empirical estimator for the F-madogram can take a value above the theoretical range of $\tfrac{1}{6}$, shown with the dotted line.}
     \label{fig:hex_plot}
\end{figure}

While having the hierarchy of partitions is useful, often a single realisation of the clustering is desired. In this instance, it is important to consider how cut heights in F-madogram space translate to euclidean space. Figure \ref{fig:hex_plot}, shows the a plot of euclidean distance against the F-madogram distance for all pairs of stations in Southwest Western Australia.  At low cut heights, the F-madogram distance changes rapidly relative to very small changes in euclidean distance. At high cut heights, large changes in euclidean distance are observed for small changes in F-madogram distance. Therefore there is a range of moderate cut heights that will translate into meaningful partitions of our stations in terms of extremal dependence in euclidean space. For this Figure, suitable cut heights might be between 0.1 and 0.15. The cut height should therefore be chosen based on the desired application and the desired strength of extremal dependence.

\subsection{Visualisation of Full Dependence} 

To understand how our regionalisation is related to the full extremal dependence, we have taken the additional step of fitting a Smith model. The full extremal dependence of each region can then be visualised using elliptical level curves. \\

\begin{figure}[!htb]
	\center
		\includegraphics[width=0.5\textwidth]{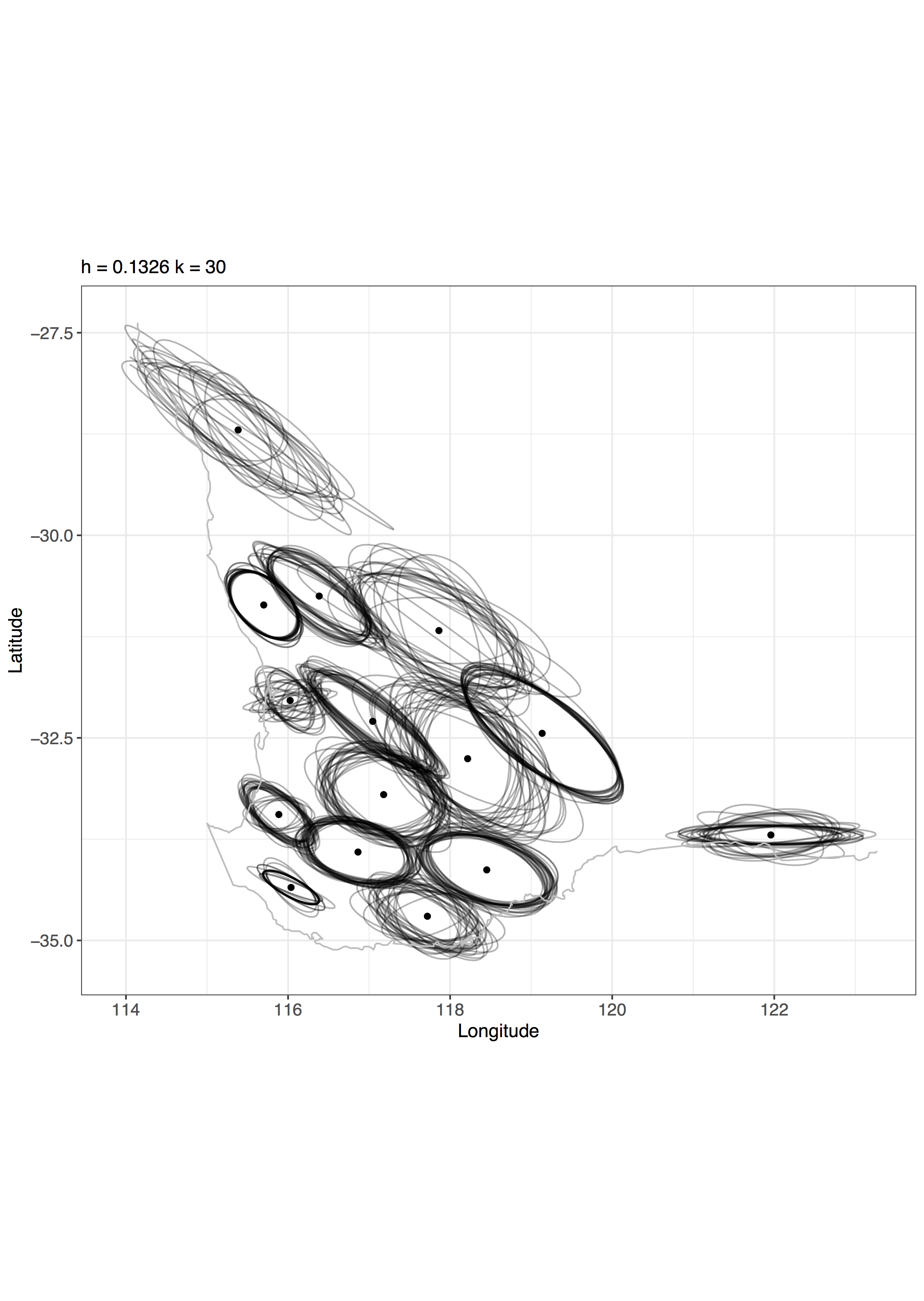}
		\caption{For a regionalisation generated with a cut height of $\sim$ 0.13, the full dependence is visualised using elliptical level curves. The black points show the median of the stations in that region and elliptical centres.}
     \label{fig:smith_SWWA}
\end{figure}

An example of the elliptical level curves is shown in Figure \ref{fig:smith_SWWA}. We observe that the ellipses have optimally partitioned the domain, as no further ellipses could be added or removed. To be confident in this conclusion we have bootstrap sampled the stations and repeated the fitting to visualise the uncertainty in our dependence parameters. We found fitting max-stable models of this type to be useful in deciding the number of clusters and to identify which regions can reasonably be modelled using the same dependence structure. We also develop an intuition for which covariates would be necessary if a non-stationary dependence structure was used. 

\subsection{Physical Interpretation}

\begin{figure*}[!h]
	\center
		\includegraphics[width= 0.8\textwidth]{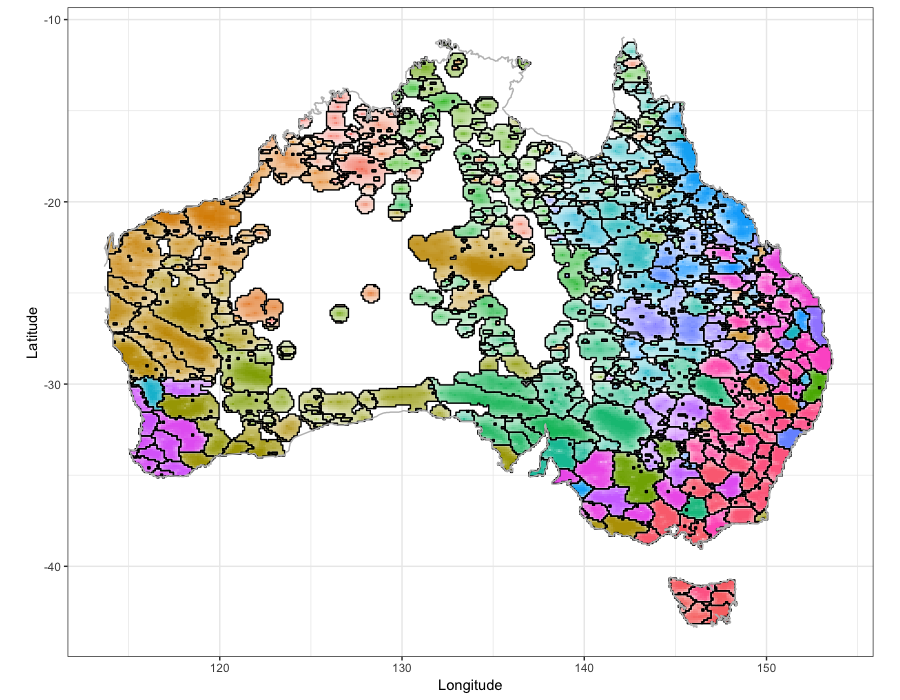}
		\caption{A regionalisation generated with a cut height of $\sim$ 0.13. Here the colours serve only to distinguish between regions.}
     \label{fig:all_aus}
\end{figure*}

\begin{figure}[h]
	\center
		\includegraphics[width = 0.6\textwidth]{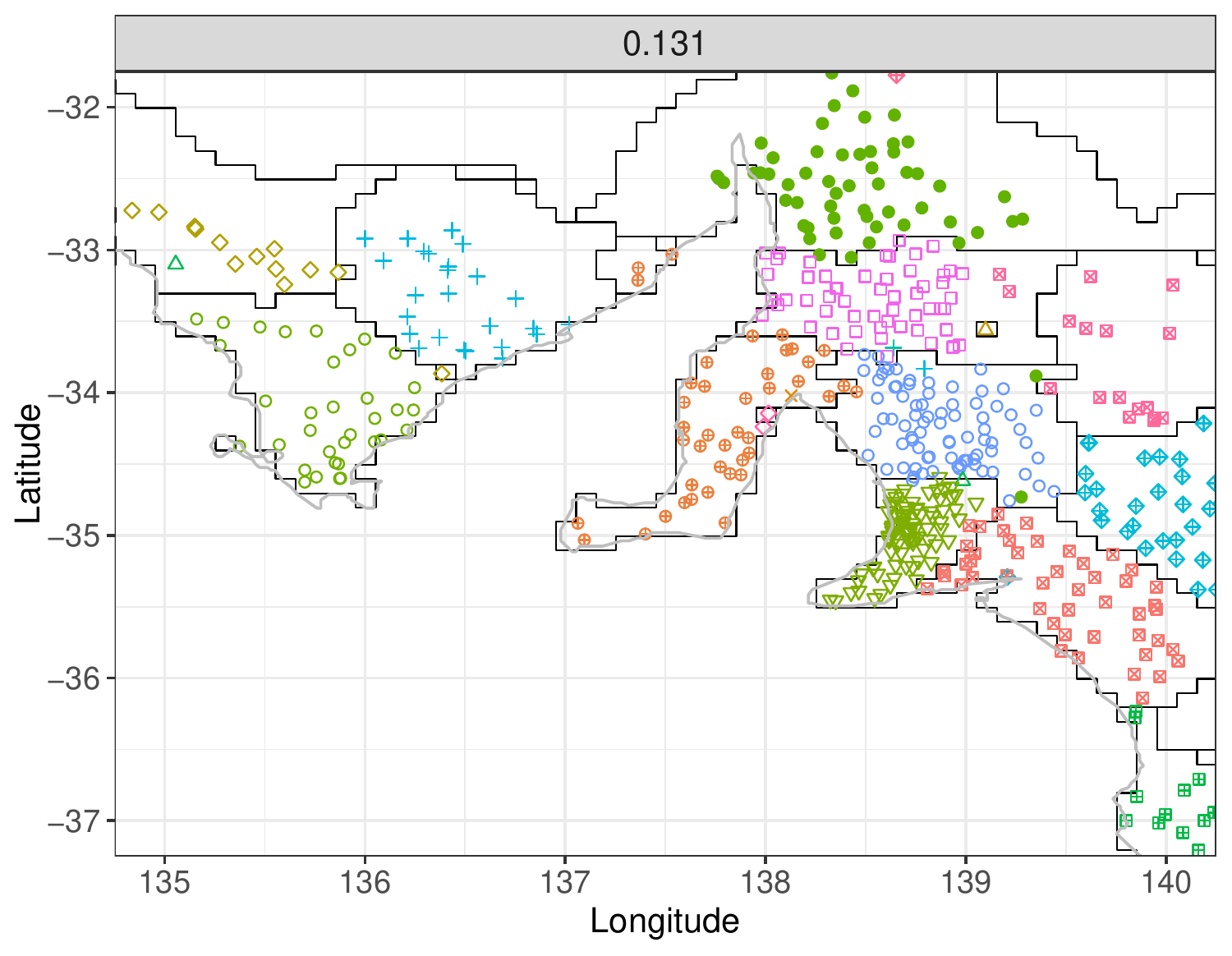}
		\caption{An example demonstrating the clusters respects stations that are separated by water. The black lines show the regions and the shape and colour of the points indicate which stations were clustered similarly.}
     \label{fig:sea_points}
\end{figure}

\begin{figure}[h]
	\center
		\includegraphics[width = 0.6\textwidth]{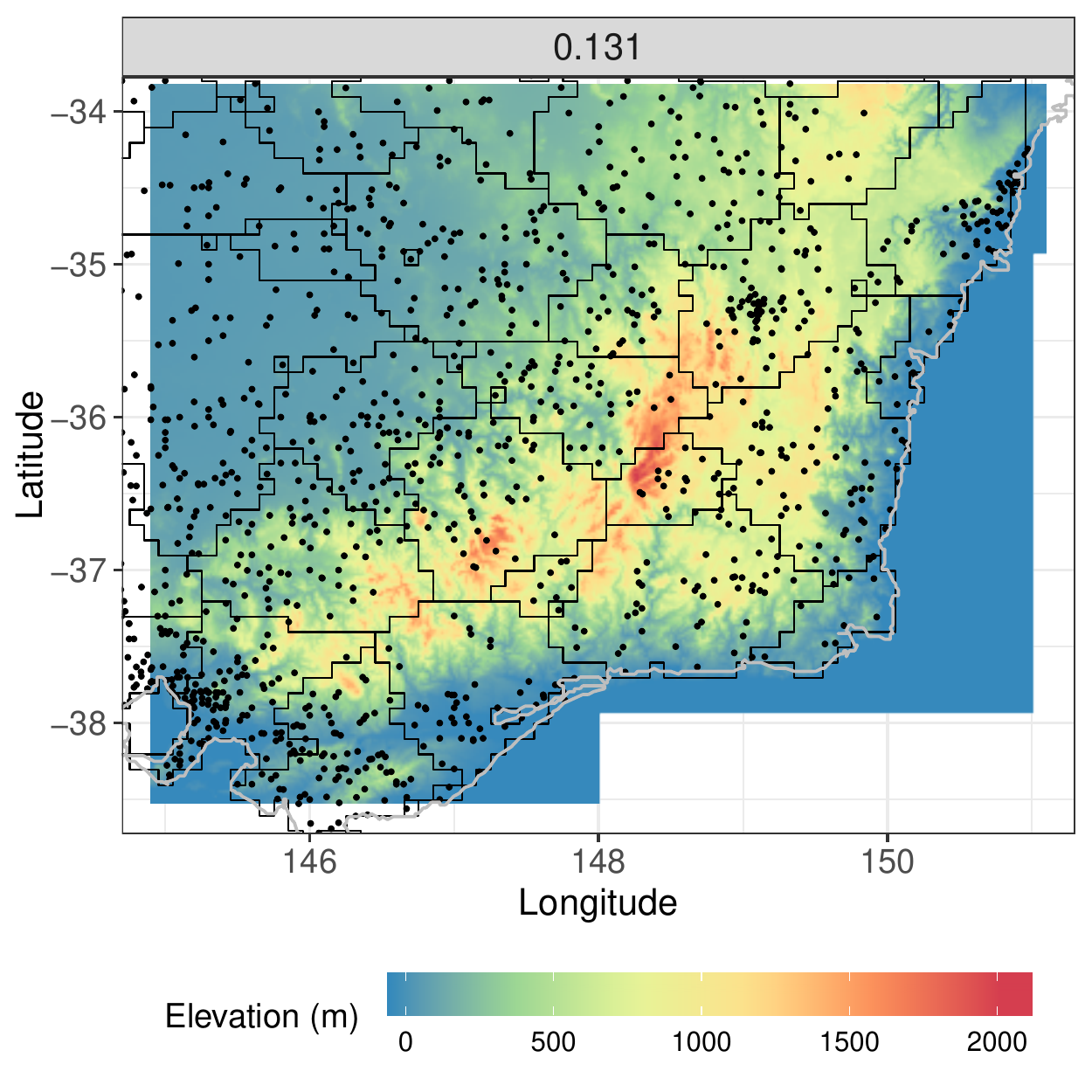}
		\caption{An example demonstrating that the clustering respects the location of the Great Dividing Range, a mountain range in Australia. Here black lines show the regions and stations are shown as black points for reference.}
     \label{fig:sea_gdr}
\end{figure} 

The example of Southwest Western Australia has served to highlight  different aspects that need to be considered when producing a regionalisation. For this same cut height, we have shown the regionalisation for the whole of Australia in Figure \ref{fig:all_aus}. Note we did not attempt to classify locations that were far from station locations. \\

We would like to draw attention to specific aspects within this figure where the regionalisation method has performed well. Figure \ref{fig:sea_points} shows examples where the clustering respects that stations are geographically separated by water. Figure \ref{fig:sea_gdr} shows how the regionalisation performs relative to orography.\\

Orographic features are known to strongly influence rainfall. In Australia there is a mountain range that runs up the Eastern Australia coast. We see this orographic feature respected in the Figure \ref{fig:sea_gdr}. There is a clear differentiation between clusters located on the coastal side of the range and those inland. This again reflects differences in the drivers between extremes in coastal areas compared with inland areas \citep{risbey2009remote}. 


\section{Limitations}

\subsection{Dry regions}

The F-madogram distance has interpretation in terms of the partial dependence of extremes provided the extreme value theory assumptions are reasonable. However for drier regions, such as parts of inland Australia and Northern Australia, where there is less rainfall, these assumptions are generally invalid \citep{min2013influence}. As a consequence the clustering will lack the  interpretation in terms of extremal dependence, impacting the related visualisation of the clustering in euclidean space, and we observe this in Figure \ref{fig:all_aus}. Therefore stations located in dry regions should be considered critically in this kind of analysis. 

\subsection{Partial Dependence}

For a given regionalisation, it is tempting to assume that within each region we can assume a fixed dependence structure in our statistical models. However, as acknowledged, the F-madogram is only a measure of partial extremal dependence, not the full extremal dependence. For regions that encompass orography, a single dependence structure is unlikely to be appropriate \citep[eg.][]{oesting2017statistical, huser2016non}. \\

\begin{figure}[h]
	\center
		\includegraphics[width=0.45\textwidth]{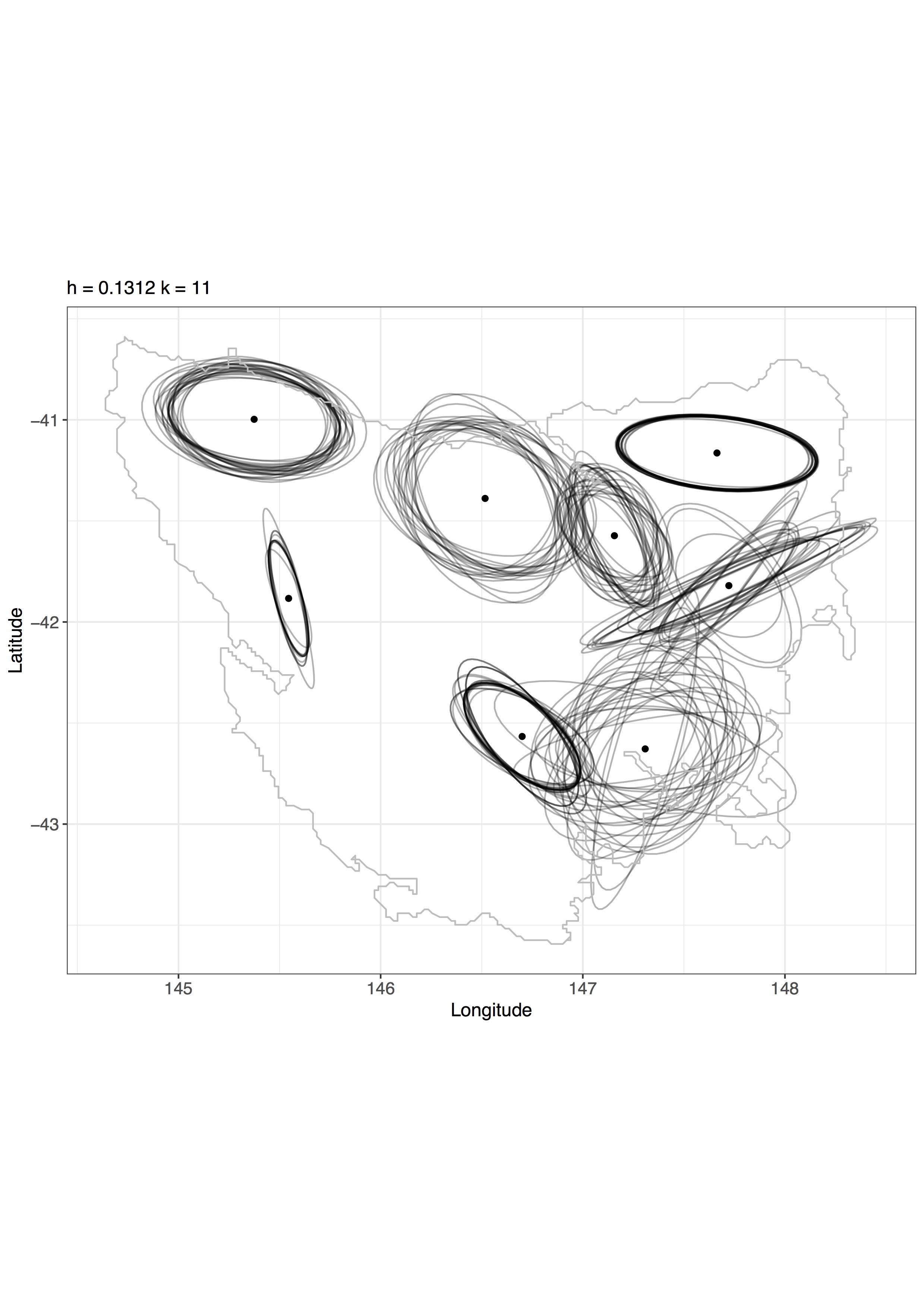}
		\includegraphics[width=0.45\textwidth]{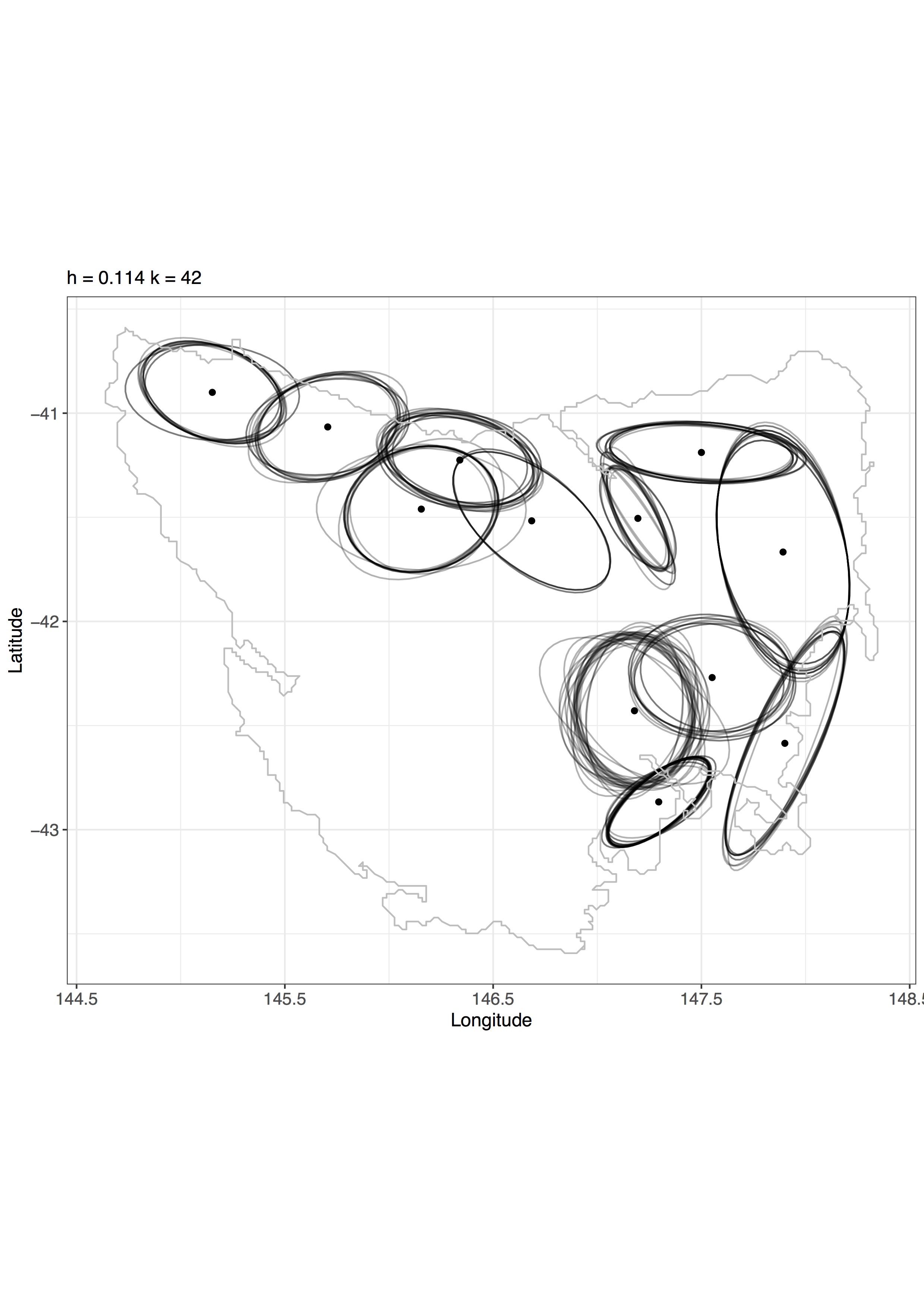}\caption{Visualisation of the full dependence for two different regionalisation. The left figure was generated with a cut height of $\sim$ 0.13 and the right figure with a cut height of $\sim$ 0.11.}
     \label{fig:smith_TAS}
\end{figure}

We observe this to be the case for regions in Tasmania, Figure \ref{fig:smith_TAS}. At a higher cut height ($\sim$ 0.13), where partial dependence is weaker within clusters, there is no consensus in the size and orientation of the ellipses for regions that encompass orography. At the lower cut height however ($\sim$ 0.11), where the dependence within clusters is stronger, there is consensus in our fitted models. Cut heights therefore need to be chosen with respect to the given application. \\

\begin{figure}[!h]
	\center
		\includegraphics[width=0.5\textwidth]{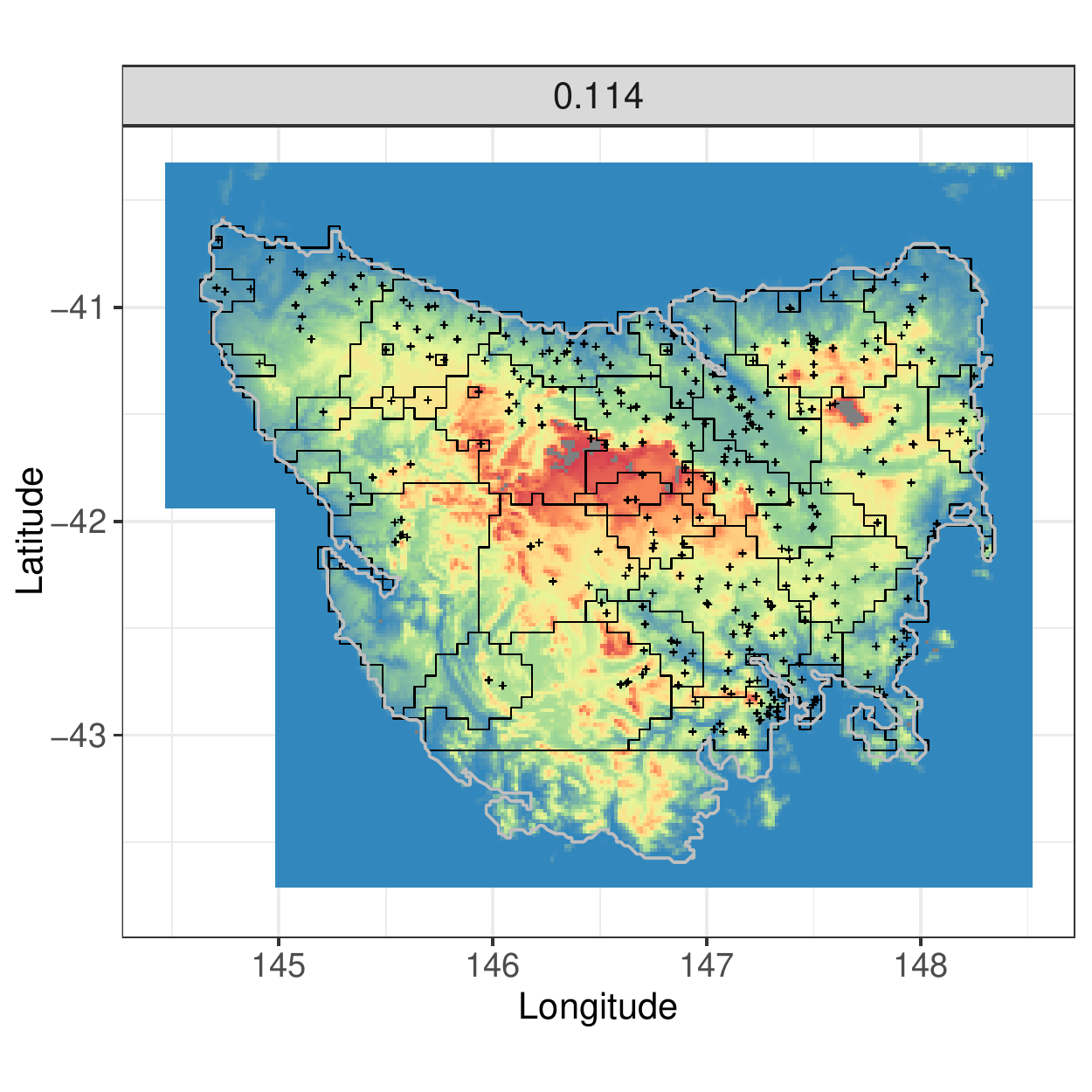}
		\caption{The regionalisation of Tasmania at a cut heigh of $\sim$ 0.11 overlaid on an elevation map.}
     \label{fig:elev_TAS}
\end{figure}

We also note that the location of cluster boundaries at the lower cut height better respects the orography, Figure \ref{fig:elev_TAS}. This regionalisation is consistent with assertions by \citet{grose2010climate} that many small regions are needed for rainfall compared with the East-West split advocated within the National Resource Management (NRM) clusters \citep{bureau2015climate}.


\section{Conclusions}

Using hierarchical clustering with the F-madogram distance, we have created a regionalisation based on the dependence of rainfall extremes. The clustering produced coherent partitions in euclidean space. This was despite using only the observed, daily annual maxima. Additionally the regions generated from the clusters are broadly consistent with our understanding of climate and topographic features \citep{stern2000objective, risbey2009remote}. Given its simplicity, the regionalisation method we have presented is therefore very powerful.\\

Climate scientists, hydrologists and other researchers can use these regionalisations to improve their understanding about the behaviour of rainfall extremes. The size and shape of the regions provides information about the range of dependence and direction of anisotropy. Also, we can produce different regionalisations for different cut heights, where different cut heights, correspond to different levels of regional detail relative to the desired strength of extremal dependence. \\

In addition to presenting the regionalisation, we highlighted key methodological considerations when using the F-madogram distance for clustering. The F-madogram distance can produce spurious clustering, depending on the underlying station network and the clustering method used. For clustering algorithms that are sensitive to point density this is of particular concern. Therefore for our application, $K$-medoids was completely unsuitable. This motivated using hierarchical clustering.\\

In general, we would advocate for using hierarchical clustering over $K$-medoids for two reasons. The agglomerative implementation of hierarchical clustering improves cluster cohesion. Additionally, the ordered partitions have an interpretation in terms of dependence strength.\\

To understand the partitions relative to the full extremal dependence, we took the additional step of fitting a max-stable models. As the dependence structure of our chosen max-stable model was Gaussian, we visualised the range of dependence and direction of anisotropy using elliptical level curves. For our regionalisations, we observed that there are many and varied dependence structures for rainfall extremes in Australia. Even for small regions we found that assuming a single dependence structure was not always suitable, but it depended on topographic features and cut height chosen. \\

There are many future directions of this research. Our approach to producing regionalisations can be used to consider different maxima, such as monthly maxima, or different variables, such as temperature. Additionally, here we have also assumed stationarity, but we are curious as to how the dependence of rainfall extremes may vary temporally, such as under different large scale climate drivers \citep{saunders2017spatial, min2013influence} or under a changing climate \citep{westra2013global, alexander2017historical}. We would be interested in comparing regionalisation from this method under different time periods \citep{bador2015spatial} and comparing regionalisations generated using observations to those from gridded data sets \citep{jones2009high}. \\

Our future goal for this research is to use the insights to model rainfall extremes on a continental scale, and to understand the impacts across large geographical distances. The regionalisations created can be used to help inform covariate selection and model selection for max-stable processes with non-stationary dependence \citep{huser2016non}. When we started this research, this goal was aspirational. However, given the knowledge generated about the behaviour and dependence of rainfall extremes in Australia, this is now a very tangible direction for future research. 

\section*{Acknowledgments}

Kate Saunders would like to thank Peter Taylor for his support and guidance throughout the course of her Ph.D, during which this research was undertaken. She would also like to thank Phillippe Naveau for his helpful suggestions and guidance during the onset of this work.\\


\bibliographystyle{jrss} 

\bibliography{bibliography_paper}

\end{document}